\documentclass[aps,prb,twocolumn,groupedaddress]{revtex4}

\usepackage{graphicx}

\begin{document}


\title{Disorder Effect on the Vortex Pinning by the Cooling-Process Control in the Organic Superconductor $\kappa$-(BEDT-TTF)$_2$Cu[N(CN)$_2$]Br}


\author{N. Yoneyama}
\author{T. Sasaki}
\author{N. Nishizaki}
\author{N. Kobayashi}
\affiliation{Institute for Materials Research, Tohoku University, Sendai 980-8577, Japan}



\date{\today}

\begin{abstract}
We investigate the influence of disorders in terminal ethylene groups of 
BEDT-TTF molecules (ethylene-disorders) on the vortex pinning of the 
organic superconductor $\kappa$-(BEDT-TTF)$_2$Cu[N(CN)$_2$]Br.
Magnetization measurements are performed under different cooling-processes.
The second peak in the magnetization hysteresis curve is observed for 
all samples studied, and the hysteresis width of the magnetization 
becomes narrower by cooling faster.
In contradiction to the simple pinning effect of disorder, 
this result shows the suppression of the vortex pinning force by 
introducing more ethylene-disorders.
The ethylene-disorder domain model is proposed for explaining the 
observed result.
In the case of the system containing a moderate number of the 
ethylene-disorders, the disordered molecules form a domain structure and  
it works as an effective pinning site.
On the contrary, an excess number of the ethylene-disorders may weaken the effect of 
the domain structure, which results in the less effective pinning force 
on the vortices.
\end{abstract}

\pacs{74.70.Kn}

\maketitle


\section{Introduction}
BEDT-TTF based organic superconductors, where
BEDT-TTF is bis(ethylene\-dithio)\-tetrathiafulvalene, have the
layered structure and the two-dimensional electronic 
properties.\cite{IshiguroYamajiSaito} 
As a feature of molecular-based compounds, the internal degree of 
freedom in the BEDT-TTF molecule can give a positional disorder.
\begin{figure}
\includegraphics[viewport=0cm 0cm 10cm 6cm,clip,width=0.9\linewidth]{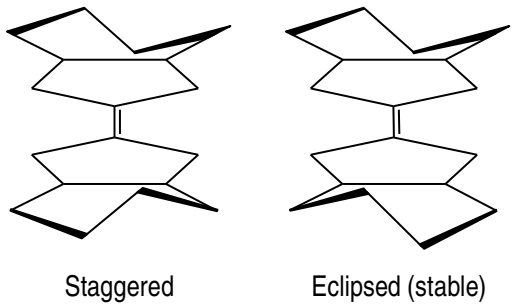}
\caption{Schematic view of conformational disorder in BEDT-TTF.}
 \label{Fig:ET}
\end{figure}

The terminal ethylene groups of the molecule can be thermally 
excited between two conformations, ``eclipsed'' and ``staggered'' (Fig.\ 
\ref{Fig:ET}).
In the present compound, $\kappa$-(BEDT-TTF)$_2$\-Cu[N(CN)]$_2$Br, one 
of the ethylene group in a molecule is known to be disordered and the
other ordered at room temperature,\cite{231} where the eclipsed 
conformation is stable. 
Such a thermal excitation is observed in the narrowing effect of the 
$^{13}$C-NMR line width above $\sim $150 K.\cite{226}
The conformational disorder (ethylene-disorder) is frozen by 
quenching, and the introduction of the ethylene-disorders slightly 
affect the electronic state.
The resistivity below about 80 K depends on cooling-rate,\cite{146} and 
in addition the superconducting transition temperature $T_c$ is also 
reduced by cooling faster.\cite{227}
The heat capacity measurement suggested that a glass transition
is caused by freezing of the motion in the ethylene group around
100 K.\cite{170}
Tanatar \textit{et al.} proposed the $T$-$P$ phase diagram on ordering
of the ethylene groups.\cite{201}
Stalcup \textit{et al.} reported that 
a time-dependent resistivity is correlated with $T_c$ and the amplitude of 
the Shubnikov-de Haas oscillation in terms of the 
ethylene-disorders.\cite{278}
However, none of the reports have clarified the spatial distribution of 
the ethylene-disorders at all.

On the superconducting mixed states of the $\kappa$-phase BEDT-TTF 
organic superconductors, following features have been observed;
the wide reversible magnetization regime,\cite{156} the second
magnetization peaks,\cite{160,155,255,162,281} and the
first-order phase transition between the vortex solid and liquid
states.\cite{158,152,161,890}
Taniguchi and Kanoda reported the cooling-rate dependence of the 
magnetization curves,\cite{255} where the hysteretic magnetization width 
becomes narrow by cooling fast. 
One may consider that cooling fast introduces a much larger number of 
the ethylene-disorders, and leads to much stronger vortex
pinning on the basis of a concept of the simple pinning effect by disorder.
Nevertheless, this expectation does not agree with their experiments. 
The nature of this controversial result is unclear.

In this paper, we report detailed measurements of the hysteretic 
magnetizations under the control of cooling-process.
We clarify the effect of the ethylene-disorders on the vortex pinning, 
and discuss the spatial distribution of the ethylene-disorders
for explaining the observed results.
We propose that domain structures originated from the 
ethylene-disorders work as a pinning site.

\section{Experimentals}
Single crystals were synthesized by means of electrocrystallization.
All the measurements of the magnetization were performed
using a SQUID magnetometer (Quantum Design, MPMS-5) for a 
single crystal with the dimension of $1.8 \times 1.6 \times
0.3 \mathrm{mm}^3$.
No grease was used to avoid strain to the sample.
The magnetic-field was applied perpendicular to the conduction plane.
The cooling-process was controlled from room temperature to 15 K as follows;
(1) 0.17 K/min with 30 h anneal at 70 K (slow-cooled), (2)
15 K/min (rapid-cooled), and (3) 100 K/min (quenched).
After all the sequential measurements of the quenched sample, the data
taken in the slow-cooled process again were completely reproduced,
suggesting no deterioration of the sample by quenching, such as micro
cracks.
We carried out the similar measurements for other two samples, which
indicate the same results with the present report.

\section{Results}
\begin{figure}
\includegraphics[viewport=0cm 0cm 13cm 13cm,clip,width=0.9\linewidth]{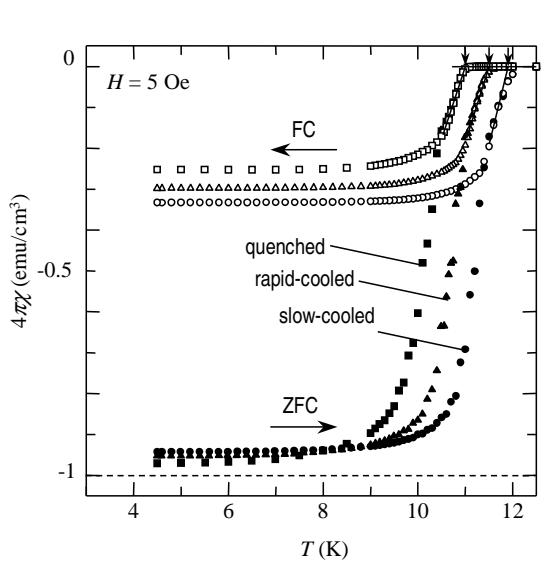}
\caption{Temperature dependence of the static magnetic susceptibility at
 5 Oe.
 The closed and open symbols indicate the shielding curve measured under
a zero-field-cooling condition (ZFC) and the Meissner curve under
a field-cooling (FC), respectively, with overlaid for the slow-cooled
(circles), rapid-cooled (triangles), and quenched processes (squares).
The broken line shows the full-Meissner volume.}
 \label{Fig:kai}
\end{figure}

Figure \ref{Fig:kai} shows the temperature dependence of the static
magnetic susceptibility at 5 Oe.
After subtracting the contribution of the core diamagnetization
($-4.8 \times 10^{-4}$ emu/mol), the demagnetization factor was
corrected using an ellipsoidal approximation.
$T_c$'s defined as an intercept of the extrapolated lines of the
normal and reversible susceptibilities are 11.9, 11.5, and 11.0 K in the
slow-cooled, rapid-cooled, and quenched process, respectively (short 
arrows in Fig. \ref{Fig:kai}).
The suppression of $T_c$ by cooling faster is quantitatively
consistent with the previous report.\cite{227,785}
The magnetizations under a zero-field-cooled (ZFC) condition 
(shielding curves) show almost the full Meissner volume and 
the negligible cooling-rate dependence within the accuracy of our 
measurement.
This result demonstrates that the superconducting volume fraction is 
almost unchanged by cooling-process, whereas Pinteri\'{c} \textit{et al.} 
reported a small suppression of the superconducting volume fraction by 
faster-cooling.\cite{785}
On the other hand, the magnetic susceptibilities under a field-cooled (FC)
condition (Meissner curves) are reduced in comparison with those under 
the ZFC condition (open symbols in Fig.\ \ref{Fig:kai}).
The volume fractions are slightly suppressed
with increasing cooling-rate.
At first sight, the suppression might be attributed as reinforcement of 
vortex pinning by cooling fast, because smaller exclusion of 
magnetic-field generally reflects stronger pinning.
However, as shown later, the experimental results in higher 
magnetic-fields are inconsistent with this explanation. 
We so suggest the other scenario as follows.
It should be noted that the magnetic-field of 5 Oe for the susceptibility
measurements is not sufficiently smaller than 
$H_{c1}(0)$ = 19 G.\cite{180} 
$H_{c1}(0)$ seems to decrease with $T_c$ by cooling faster, and 
therefore the effective field $H/H_{c1}(0)$ is increased.
This makes the flux exclusion more incomplete, leading to
the further suppression of the FC superconducting volume by cooling faster.

\begin{figure}
\includegraphics[viewport=0cm 0cm 13cm 15cm,clip,width=0.9\linewidth]{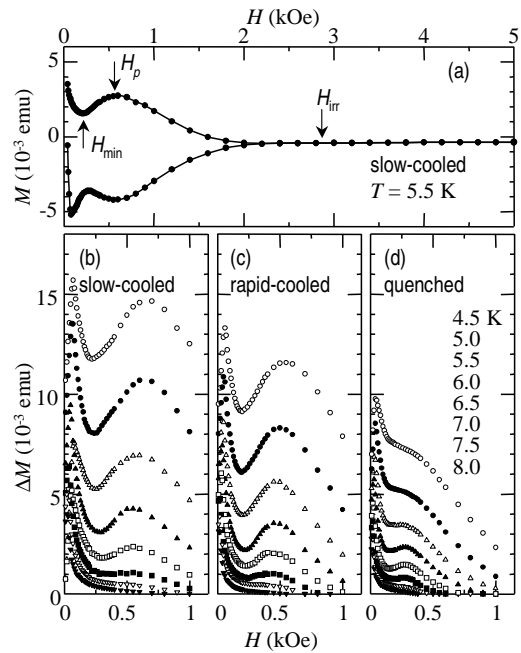}
\caption{Magnetic-field dependence of (a) the magnetization at 5.5 K,
 and the magnetization hysteresis width for cooling-rates of (b)
 slow-cooled, (c) rapid-cooled, and (d) quenched condition,
 respectively.}
\label{Fig:MH}
\end{figure}

Figure \ref{Fig:MH}(a) shows the magnetic-field dependence of 
the magnetization at 5.5 K.
A typical second peak is found around 500 Oe.
As displayed in the magnetization hysteresis width $\Delta M$
(Fig.\ \ref{Fig:MH}(b)--(d)), the second peak is observed below about 8 K
for all the cooling-process. 
The peak height seems to decrease with increasing cooling-rate as
reported previously.\cite{255}
Strictly speaking, to depict the cooling-rate dependence of $\Delta M$, a 
comparison of the measured $\Delta M$'s should 
be done at a fixed temperature $t$ (= $T/T_c$), since cooling-rate
changes $T_c$.

\begin{figure}
\includegraphics[viewport=0cm 0cm 13cm 14cm,clip,width=0.9\linewidth]{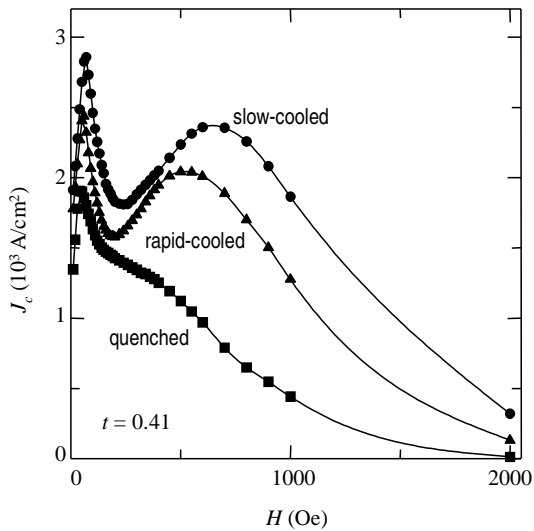}
\caption{Magnetic-field dependence of the critical current density at
$t=0.41$ estimated from $\Delta M(H)$ on the basis of a standard Bean's 
model\cite{Tinkham} for a square-shape sample\cite{596}.} 
 \label{Fig:jc}
\end{figure}

Figure \ref{Fig:jc} shows the magnetic-field dependence of the
critical current density $J_c$ at $t$ = 0.41.
The estimation of $J_c$ from $\Delta M$ was carried out through a
standard Bean's model\cite{Tinkham} for a square-shape sample\cite{596}.
The data at 4.5 K are used for the quenched sample, and
the corresponding data at 4.7 K (rapid-cooled) and 4.9 K (slow-cooled)
are calculated from a linear interpolation with the data at both 4.5 and
5.0 K, respectively.
Even after the subtraction of the effect on the reduced $T_c$,
the critical current density clearly decreases with increasing cooling-rate.
This indicates the suppression of the effective vortex pinning by cooling fast.
Because rapid cooling of samples is considered to induce frozen 
ethylene-disorders, one may expect the increase of vortex pinning.
However, in the present vortex system, the effective pinning force 
becomes weak with increasing ethylene-disorder.
This result is apparently inconsistent with a concept of the simple 
pinning driven by disorder.

\begin{figure}
\includegraphics[viewport=0cm 0cm 13cm 17cm,clip,width=0.9\linewidth]{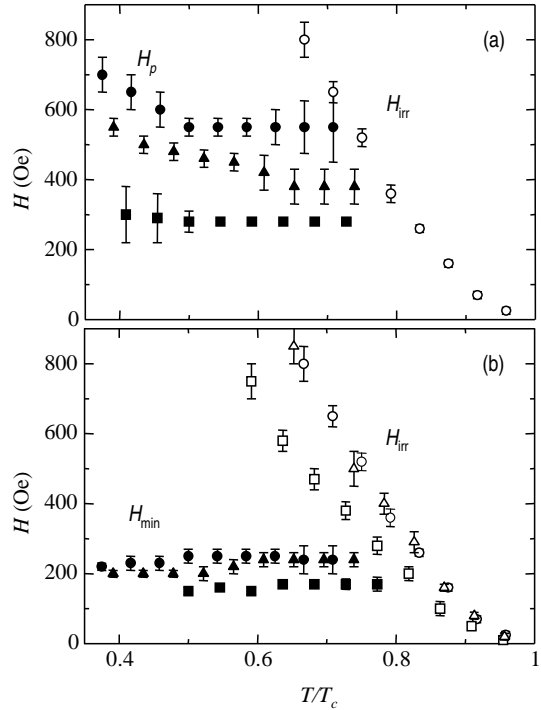}
\caption{Characteristic magnetic-fields (closed symbols) as a function
 of temperature with overlaid for the slow-cooled (circles), rapid-cooled
 (triangles), and quenched (squares): (a) $H_p$ and (b) $H_{\text{min}}$.
 $H_{\text{irr}}$ (open symbols) is also plotted.
 }
 \label{Fig:PD}
\end{figure}

Here we characterize the magnetic-field of the second peak.
Contrary to the peak effect of Bi$_2$Sr$_2$CaCu$_2$O$_8$ 
(Bi2212),\cite{Bi2212} 
the peak structure in the present superconductor is broad.
This makes it difficult to determine a characteristic magnetic-field of 
the second peak  without any ambiguity.
We here define the peak position using two magnetic-fields:
one is the magnetic-field of the minimum point between the central
and second peaks ($H_{\text{min}}$), and the other is the maximum point
for the peak ($H_p$).
Each magnetic-field is plotted in the $T$-$H$ phase diagrams with
overlaid for each cooling-process, as shown in Fig.\ \ref{Fig:PD}.
Additionally plotted irreversibility fields ($H_{\text{irr}}$, open
symbols) are defined using a criterion of $J_c=2.0$ A/cm$^2$
(corresponding to $\Delta M=1.0\times 10^{-5}$ emu).
As a function of temperature, $H_p$ increases with reducing temperature
for all the cooling-process (Fig.\ \ref{Fig:PD}(a)).
On the contrary, $H_{\text{min}}$ shows the different behavior as
displayed in Fig.\ \ref{Fig:PD}(b), where the $H_{\text{min}}$ lines are
piled around 200 Oe.

\section{Discussion}
We first discuss the origin of the second peak effect.
The peak effect of the present compound has been attributed to the
dimensional crossover of the vortex system.\cite{160,155,255}
A vortex line changes to two-dimensional (2D) pancake vortices with the
decrease of the interlayer Josephson coupling above
the crossover field $B_{\text{cr}} \approx \phi_0/(\gamma d)^2$, where
$\phi_0$ is the magnetic flux quantum, $\gamma$ the anisotropy factor,
and $d$ the interlayer distance.
With the values of $\gamma=$ 130 -- 180 \cite{20,257} and
$d=15$ \AA, one obtains $B_{\text{cr}}\approx$ 280 -- 540 G.
The estimated $B_{\text{cr}}$ roughly corresponds to the second peak.
It is almost the same as $H_{\text{min}}$ and $H_p$, 
and thus the dimensional crossover scenario as the second peak origin is 
applicable to this case.
Although the broad shape of the second peak seems to be inconsistent with 
this interpretation, the broadness may come from the averaging of the 
peak structure over the bulk sample in which the local magnetic-field 
has a distribution.
A more detailed experimental investigation would be needed to check this 
point.
$B_{\text{cr}}$ is generally independent of temperature, and moreover
in the present system, cooling-rate independence is expected as well, because
the anisotropy factor $\gamma$ is likely to be unchanged by cooling-rate.
Actually, even in the deuterated BEDT-TTF based-salt, which has 
dramatically different electronic properties from the present 
(non-deuterated) salt,\cite{203} the same value of $\gamma$ has been 
reported.\cite{138}
While $H_p$ decreases with increasing both temperature and cooling-rate,
$H_{\text{min}}$ is almost independent of both of them.
The behavior of $H_p$ cannot be explained by a simple dimensional crossover 
model.
On the contrary, $H_{\text{min}}$ seems to be appropriate as $B_{\text{cr}}$.
If $B_{\text{cr}} \sim H_{\text{min}}$, the second peak is qualitatively 
explained as follows:
In $H > \sim H_{\text{min}}$, two-dimensional pancakes appear, 
and the pinning force for the pancake vortices is effectively optimized 
around $H_p$.

Besides dimensional crossover, several models have been proposed as the 
origin of the second peak effect: (1) matching effects,\cite{matching}
(2) the magnetic-field-dependent relaxation-rate (collective pinning 
model),\cite{476} (3) the field-induced order-disorder transition
(Bragg-glass to vortex-glass transition), etc.
First, in (1) matching effects, the peak structure appears
around a magnetic-field in which the vortex lattice structure is 
commensurate with the defect structure.
In this model, when the number of defects increases, the magnetic-field 
of the peak maximum increases.
However in the present case, by cooling faster which may be giving 
much larger number of defects, $H_p$ shifts toward lower magnetic-fields.
Thus the experimental results are inconsistent with this model.
Next, the second model (2) is also excluded by the following reason;
in this model, a collective pinning of vortices appears above the peak 
maximum field, resulting in a slower relaxation-rate than that below the 
magnetic-field where a single-vortex pinning mechanism governs the system.
According to a relaxation measurement in the present compound\cite{155},
the relaxation-rate at the peak maximum field is faster than 
that at the peak minimum one, which contradicts with this model.
The third model (3) is relatively controversial.
In this model, with increasing magnetic-field, 
a first-order phase transition happens from a vortex lattice or Bragg 
glass state (ordered-solid phase) to a vortex-glass state 
(disordered-solid phase), lacking translational order.
This transition is originated from the larger pinning energy gain of the 
glass phase than that of the lattice one, which can reasonably explain 
the peak effect.
In high-$T_c$ superconductors\cite{501,394}, the order-disorder transition 
has been observed as a steep increase or a step of magnetization, but in 
the present organic superconductor, there has been no decisive evidence 
in favor of this model.
To confirm this model, it is needed to investigate, for example,
the magnetization under a rapidly fluctuating magnetic-field, 
as has been measured in Bi2212.\cite{394}

While the dimensional crossover explains the second peak effect,
the problem of the suppressed $J_c$ by cooling fast is left unsettled.
Note that the suppression is remarkably shown in $H > H_{\text{min}}$ (Fig.\ 
\ref{Fig:jc}), in which the vortex system consists of pancakes.
This implies the weakness of pinning, which will support the discussion 
on domain structures of the ethylene-disorders as a weak pinning origin, 
presented later.
In the following, to understand the observed result we solve the 
puzzling origin of the vortex pinning.

As a pinning site, intrinsic defects such as molecular defects,
dislocations, or coexist of another crystal phases are considerable. 
Actually a coexisting $\alpha$-phase in the $\kappa$-type crystal has been 
observed using scanning tunneling microscopy (STM).\cite{309}
However, these intrinsic defects, if any, are independent of cooling-rate.
As already mentioned, our experimental results indicate that the effective 
pinning force is suppressed with cooling faster, that is, with 
increasing ethylene-disorder.
Here it should be noted that the size of a BEDT-TTF molecule ($\sim 3$
\AA \ projected on the $ac$-plane) is much smaller than that of a vortex core
($2\xi_\parallel (0) \sim 50$ \AA, where $\xi_\parallel(0)$ is the
in-plane coherence length \cite{156,138}).
Thus it is difficult for an \textit{individual} ethylene-disorder to behave 
itself as a pinning site because of smallness of the molecular size.

We next describe the numbers of the ethylene-disorders.
In the present case, the quenched condition will give the largest number of 
the ethylene-disorders, because the conformational disorders at room
temperature are frozen by quenching.
The fraction of the ethylene-disorders in the quenched condition is 
suggested to be about 5--20\%.\cite{170,329}
Although there are no corresponding reports after slow-cooling,
we suppose that a slight number of the ethylene-disorders exist even in our
slow-cooled condition.
Tokumoto \textit{et al.}\cite{227} reported that $T_c$ monotonically decreases 
from 12.2 K to 11.7 K as the cooling-rate increases from 0.02 K/min up 
to 10 K/min.
$T_c$ of 11.9 K at the rate $\sim$ 0.2 K/min in their report shows a good
agreement with our slow-cooled case.
Thus $T_c$ of the present sample can furthermore increase by cooling
more slowly. 
In other words, this result suggests that there exist some 
ethylene-disorders even in the present slow-cooled condition.


\begin{figure}
\includegraphics[viewport=0cm 0cm 15cm 16cm,clip,width=0.9\linewidth]{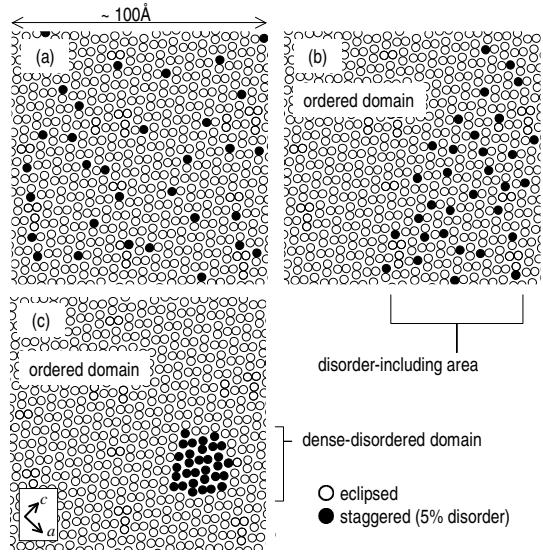}
\caption{Schematic views of relaxed states in the slow-cooled condition
 with 5\% ethylene-disorders.
 Each panel shows the molecular arrangement projected on the $ac$-plane.
 Three possible distributions of the ethylene-disorders are described: (a) 
 a random distributed state, (b) a state with ordered domains 
 originated from vortex core-scale inhomogeneity, and (c) a state consisting of 
 ordered and dense-disordered domains.
}
\label{Fig:disorder}
\end{figure}

Next, we discuss the spatial distribution of the ethylene-disorders.
At high temperatures, thermal excitation will cause a random mixing of 
the staggered and eclipsed molecules and/or these micro domains.
The quenched process freezes a higher temperature glassy state with a 
random distribution of the disordered molecules.
As the temperature is slowly lowered, the stable conformation (eclipsed)
gradually becomes dominant.
The remnant ethylene-disorders weaken the superconducting state with 
reducing $T_c$. 
Although the mechanism in the relaxation process of the 
ethylene-disorders is unclear, several relaxed states after slow-cooling
are expected.
The possible relaxed states are classified into following 
three states as shown in Fig.\ \ref{Fig:disorder}.
As an example, we here assume 5\% concentration of the 
ethylene-disorders, which will be the maximum in the slow-cooled 
condition.\cite{170}
Each panel shows the molecular arrangement projected on the 
$ac$-plane over $\sim$100 \AA,
containing about 640 BEDT-TTF and 32 disordered molecules.
The BEDT-TTF molecules are described as small circles with diameter of 
about 3 \AA, consisting of ordered (open symbols) and disordered (closed 
symbols) molecules.
Figure\ \ref{Fig:disorder}(a) shows a state with ethylene-disorders 
distributed randomly, which is similar with the quenched one except for 
the small number of the ethylene-disorders.
The second possible state shown in Fig.\ \ref{Fig:disorder}(b) features 
the inhomogeneous distribution of ethylene-disorders.
In this case, ordered domains are formed by the increase of the stable eclipsed molecules,
while remnant disorder-including areas are coexisting.
A boundary structure can be formed between the ordered domain and the 
disorder-including area,
whereas the spatial structure of the domain boundaries is unidentified.
The third state is containing domains of dense-disordered molecules (Fig.\ 
\ref{Fig:disorder}(c)).
The existence of domain structures consisting of only staggered or
eclipsed molecules were suggested by the study on the relaxation of the
resistivity.\cite{208}
Here it should be noted that the interlayer correlation of the 
disorder's distribution is now out of question, since we now treat the pinning 
of the two-dimensional pancake vortices.
Therefore our description of the ethylene-disorder's distribution is also
two-dimensional one, but this does not rule out the existence of the 
interlayer correlation of the domain structure.

Taking into account the above information on the ethylene-disorders, we 
propose the domain structures of the relaxed state shown in Fig.\ 
\ref{Fig:disorder}(b) as a vortex pinning origin. 
This relaxed state contains the characteristic boundary structure 
between the ordered domain and the disorder-including area.
The size of such a domain structure is much larger than at 
least the in-plane coherence length $\sim \xi_\parallel$ (the vortex-core-scale).
In the slow-cooled condition, the widely-ordered domains dividing the 
disorder-including areas partly are formed, resulting in the vortex-core-scale 
inhomogeneity.
Therefore, if these domain structures work as vortex pinning sites, the 
slow-cooled process gives much stronger effective pinning than the 
quenched one.

On the other hand, in the relaxed state shown in Fig. \ref{Fig:disorder}(a), 
many individual ethylene-disorders are distributed randomly, giving
a quasi-homogeneous state on the vortex-core-scale.
Therefore, the ethylene-disorders do not work as vortex pinning sites, 
whereas they give rise to many point defects. 
The quenched condition is similar with this state,
while the concentration of the ethylene-disorders is much larger.
Such an excess number of the ethylene-disorders may weaken the effect of 
the domain structure, resulting in the less effective pinning force 
on the vortices.
This is consistent with our experimental results.

One may expect that the dense-disordered domain shown in Fig.\ 
\ref{Fig:disorder}(c) gives a strong pinning origin.
However, the size of the domain, in spite of containing 5\% 
ethylene-disorders, is still smaller than the vortex core.
Of course one can describe a much larger dense-disordered domain structure in 5\% 
ethylene-disorders than that of Fig.\ \ref{Fig:disorder}(c), but then 
the total number of the domains over the sample size becomes quite a few.
So the contribution to the vortex pinning by such a dense-disordered 
domain cannot be so strong, if any.

An experimental support for the existence of a spatial inhomogeneity was 
reported by Tanatar et al\cite{145}, who observed a magnetic viscosity 
phenomena and proposed a magnetic domain structure formation in the 
antiferromagnetically ordered state in the normal state.
Unfortunately, it is unclear whether their magnetic domain structure is 
directly related to the ethylene-disorders.
While this may be a possible solution to the present problem,
further experimental studies will be needed to check this point.

A clue might lie in a direct observation of the domain structures with local
probe, such as STM, which is planning for the near future.

\section{Conclusions}
In conclusion, we reported the hysteretic magnetization measurements
to investigate the disorder effect on the vortex pinning in the
cooling-rate of 0.17, 10, and $\sim$ 100 K/min.
The negligible cooling-rate dependence of the almost full volume 
fraction in the superconducting state demonstrates that the 
superconducting volume is unchanged by cooling-process. 
The second peak is observed below about 8 K for all the cooling-process.
We adopt the dimensional crossover scenario as the origin of the second peak.
The vortex pinning in the present system is suggested to be closely related 
to the ethylene-disorders.
However, it is difficult for an individual ethylene-disorder to 
behave itself as a pinning site because of smallness of the size.
Actually by cooling faster the critical current density decreases, and so
the effective pinning force is apparently suppressed.
Thus a simple pinning by the introduced ethylene-disorders does 
not occur.

To explain the puzzling pinning origin, we proposed the 
ethylene-disorder domain model, where domain structures form a
vortex-core-scale inhomogeneous distribution of the ethylene-disorders 
working as vortex pinning sites (shown in Fig.\ \ref{Fig:disorder}(b)). 
The size of the domain structure will be much larger than at 
least the in-plain coherence length.
On the contrary, an excess number of disorders introduced by quenching 
may weaken the effect of the domain structure, which results in the less 
effective pinning force.

This research was partially supported by the Ministry of Education,
Science, Sports and Culture, Grant-in-Aid for Encouragement of Young
Scientists.

\end{document}